\documentclass[nofootinbib,prd,twocolumn,showpacs,showkeys,preprintnumbers]{revtex4-1}
\usepackage{hyperref,amssymb,amsmath,mathrsfs,bm,graphicx}
\begin{document}

\title {Axially symmetric dissipative fluids in  the quasi--static approximation}%author{L. Herrera}
\author{L. Herrera}
\email{lherrera@usal.es}
\affiliation{Escuela de F\'\i sica, Facultad de Ciencias, Universidad Central de Venezuela, Caracas, Venezuela and Instituto Universitario de F\'isica
Fundamental y Matem\'aticas, Universidad de Salamanca, Salamanca, Spain}
\author{A. Di Prisco}
\email{alicia.diprisco@ciens.ucv.ve}
\affiliation{Escuela de F\'\i sica, Facultad de Ciencias, Universidad Central de Venezuela, Caracas, Venezuela}
\author{J. Ospino}
\email{j.ospino@usal.es}
\affiliation{Departamento de Matem\'atica Aplicada and Instituto Universitario de F\'isica
Fundamental y Matem\'aticas, Universidad de Salamanca, Salamanca, Spain}
\author{J.Carot }
\email{jcarot@uib.cat}
\affiliation{Departament de  F\'{\i}sica, Universitat Illes Balears, E-07122 Palma de Mallorca, Spain}
\begin{abstract}
Using a framework based on the $1+3$ formalism we carry out a study on  axially and reflection symmetric dissipative fluids,  in the quasi--static regime. We first derive a set of invariantly defined ``velocities'',  which allow for  an inambiguous definition of the quasi--static approximation. Next we rewrite all the relevant equations in this aproximation and extract all the possible, physically relevant,  consequences ensuing the adoption of such an approximation. In particular  we show how the vorticity, the shear and the dissipative flux, may lead to situations where different kind of ``velocities'' change of sign within the fluid distribution with respect to theirs sign on the boundary surface. It is shown that states of gravitational radiation are not {\it a priori} incompatible with the quasistatic--regime. However, any such state must last for an infinite period of time, thereby diminishing its physical relevance.
\end{abstract}
\maketitle

\section{Introduction}
In the study of self--gravitating fluids we may consider three different possible regimes of evolution, namely: the static, the quasi--static and the dynamic.

In the static   case, the spacetime admits a timelike, hypersurface orthogonal, Killing vector. Thus, a coordinate system can always be choosen, such that all metric and physical variables are independent on the time like coordinate. The static case, for axially and reflection symmetric spacetimes, was studied in \cite{static}.

Next we have the full dynamic case where the system is considered to be out of equilibrium (thermal and dynamic), the general formalism to analyze this situation, for axially  and reflection symmetric spacetimes was developped in \cite{1} using a framework based on the $1+3$ formalism \cite{21cil, n1, 22cil, nin}.

In between the two regimes described above, we have the quasi--static evolution.

As  is well known, in this regime the system is assumed to evolve, although sufficiently slow, so that it can be considered to be in equilibrium at each moment (Eqs. (\ref{label1}, \ref{label2}) are satisfied, in the corresponding static case). This means that the system changes slowly, on a time scale that is very long compared to the typical
time in which the fluid reacts to a slight perturbation of hydrostatic
equilibrium. This typical time scale is called hydrostatic time scale \cite{astr1}--\cite{astr3} (sometimes this time scale is also referred to as dynamical time scale, e.g. \cite{astr3}). Thus, in this regime  the system is always very close to  hydrostatic
equilibrium and its evolution may be regarded as a sequence of equilibrium
models. 

Briefly speaking, all the relevant characteristic times of the system under consideration, should be much larger than the hydrostatic  time.

This assumption is very sensible because
the hydrostatic
time scale is very small for many phases of the life of the star \cite{astr2}.
It is of the order of $27$ minutes for the Sun, $4.5$ seconds for a white dwarf
and $10^{-4}$ seconds for a neutron star of one solar mass and $10$ Km radius.
It is well known that any of the stellar configurations mentioned above, generally (but not always), 
changes on a time scale that is very long compared to their respective
hydrostatic time scales. 

In the spherically symmetric case there exist several studies on the behaviour of fluid distributions in the quasi--satic regime (see \cite{Bo, hrw, two, wb, split1, Bec, ns} and references therein).

It is our purpose here, to make use of the framework developped in \cite{1},  to carry out a study of axially  and reflection symmetric fluids in the quasi--static regime. 

For doing that we shall need to introduce different invariantly defined ``velocities'', in terms of which the quasi--static approximation (QSA) is expressed.

As we shall see, the shear and the vorticity of the fluid, as well as the dissipative fluxes, may affect the (slow) evolution of the configuration, as to produce ``splittings'' within the fluid distribution.

It will be also shown that  in the QSA, the contributions of the gravitational radiation to the components of the super--Poynting vector do not necessarily vanish. However, as we shall show below,  it appears   that if at any given time,
the magnetic part of the Weyl tensor vanishes, then it vanishes at any other time
afterwards. Thus we should not expect gravitational radiation
from a physically meaningful system, radiating for a
finite period of time (in a given time interval) in the QSA.

\section{The metric and the source: basic definitions and notation}
We shall consider,  axially (and reflection) symmetric sources. For such a system the most general line element may be written in ``Weyl spherical coordinates'' as:

\begin{equation}
ds^2=-A^2 dt^2 + B^2 \left(dr^2
+r^2d\theta^2\right)+C^2d\phi^2+2Gd\theta dt, \label{1b}
\end{equation}
where $A, B, C, G$ are positive functions of $t$, $r$ and $\theta$. We number the coordinates $x^0=t, x^1=r, x^2= \theta, x^3=\phi$.

We shall assume that  our source is filled with an anisotropic and dissipative fluid. We are concerned with either bounded or unbounded configurations. In the former case we should further assume that the fluid is bounded by a timelike surface $\Sigma$, and junction (Darmois) conditions should be imposed there.

The energy momentum tensor may be written in the ``canonical'' form, as 
\begin{equation}
{T}_{\alpha\beta}= (\mu+P) V_\alpha V_\beta+P g _{\alpha \beta} +\Pi_{\alpha \beta}+q_\alpha V_\beta+q_\beta V_\alpha.
\label{6bis}
\end{equation}

The above is the canonical, algebraic decomposition of a second order symmetric tensor with respect to unit timelike vector, which has the standard physical meaning when $T_{\alpha \beta}$ is the energy-momentum tensor describing some energy distribution, and $V^\mu$ the four-velocity assigned by certain observer.

With the above definitions it is clear that $\mu$ is the energy
density (the eigenvalue of $T_{\alpha\beta}$ for eigenvector $V^\alpha$), $q_\alpha$ is the  heat flux, whereas  $P$ is the isotropic pressure, and $\Pi_{\alpha \beta}$ is the anisotropic tensor. We emphasize that we are considering an Eckart frame  where fluid elements are at rest.

Since we choose the fluid to be comoving in our coordinates, then
\begin{equation}
V^\alpha =\left(\frac{1}{A},0,0,0\right); \quad  V_\alpha=\left(-A,0,\frac{G}{A},0\right).
\label{m1}
\end{equation}
Next, let us  introduce the unit, spacelike vectors $\bold K, \bold L$, $\bold S$, with components
\begin{equation}
K_\alpha=(0,B,0,0); \quad  L_\alpha=\left(0,0,\frac{\sqrt{A^2B^2r^2+G^2}}{A},0\right),
\label{7}
\end{equation}
\begin{equation}
 L^\alpha=\left(\frac{G}{A\sqrt{A^2B^2r^2+G^2}},0,\frac{A}{\sqrt{A^2B^2r^2+G^2}},0\right),
\label{3n}
\end{equation}
\begin{equation}
 S_\alpha=(0,0,0,C),
\label{3nb}
\end{equation}
satisfying  the following relations:
\begin{equation}
V_{\alpha} V^{\alpha}=-K^{\alpha} K_{\alpha}=-L^{\alpha} L_{\alpha}=-S^{\alpha} S_{\alpha}=-1,
\label{4n}
\end{equation}
\begin{equation}
V_{\alpha} K^{\alpha}=V^{\alpha} L_{\alpha}=V^{\alpha} S_{\alpha}=K^{\alpha} L_{\alpha}=K^{\alpha} S_{\alpha}=S^{\alpha} L_{\alpha}=0.
\label{5n}
\end{equation}
The unitary vectors $V^\alpha, L^\alpha, S^\alpha, K^\alpha$ form a canonical  orthonormal tetrad (say  $e^{(a)}_\alpha$), such that  $$e^{(0)}_\alpha=V_\alpha,\quad e^{(1)}_\alpha=K_\alpha,\quad
e^{(2)}_\alpha=L_\alpha,\quad e^{(3)}_\alpha=S_\alpha$$ with $a=0,\,1,\,2,\,3$ (latin indices labeling different vectors of the tetrad). The  dual vector tetrad $e_{(a)}^\alpha$  is easily computed from the condition 

$$ \eta_{(a)(b)}= g_{\alpha\beta} e_{(a)}^\alpha e_{(b)}^\beta,$$
where $ \eta_{(a)(b)}$ denotes the Minkowski spacetime metric

The anisotropic tensor  may be  expressed through three scalar functions defined as (see \cite{1}):

\begin{eqnarray}
 \Pi _{KL}=K^\alpha L^\beta T_{\alpha \beta} 
, \quad , \label{7P}
\end{eqnarray}

\begin{equation}
\Pi_I=(2K^{\alpha} K^{\beta} -L^{\alpha} L^{\beta}-S^{\alpha} S^{\beta}) T_{\alpha \beta},
\label{2n}
\end{equation}
\begin{equation}
\Pi_{II}=(2L^{\alpha} L^{\beta} -S^{\alpha} S^{\beta}-K^{\alpha} K^{\beta}) T_{\alpha \beta}.
\label{2nbis}
\end{equation}

This specific choice of  these scalars is justified by the fact, that  the relevant equations used  to carry out this study,  become more compact and easier to handle, when expressed in terms of  them.

Finally, we may write the heat flux vector in terms of two scalar functions:
\begin{equation}
q_\mu=q_IK_\mu+q_{II} L_\mu
\label{qn1}
\end{equation}
or, in coordinate components (see \cite{1})

\begin{equation}
q^\mu=\left(\frac{q_{II} G}{A \sqrt{A^2B^2r^2+G^2}}, \frac{q_I}{B}, \frac{Aq_{II}}{\sqrt{A^2B^2r^2+G^2}}, 0\right),\label{q}
\end{equation}
\begin{equation}
 q_\mu=\left(0, B q_I, \frac{\sqrt{A^2B^2r^2+G^2}q_{II}}{A}, 0\right).
\label{qn}
\end{equation}
Of course, all the above quantities depend,  in general, on $t, r, \theta$.

The kinematical variables play an important role in the description of  a self--gravitating fluid. Here, besides the four acceleration, the expansion scalar and the shear tensor, we have a component of vorticity.

Thus we obtain   (see  \cite{1}).

For the four acceleration
\begin{equation}
a_\alpha=V^\beta V_{\alpha;\beta}=a_I K_\alpha+a_{II}L_\alpha,
\label{a1n}
\end{equation}
with
\begin{equation}
a_I= \frac {A^\prime }{AB };\quad a_{II}=\frac{A}{\sqrt{r^2A^2B^2+G^2}}\left[\frac {A_{,\theta}}{A}+\frac {G}{A^2}\left(\frac{\dot G}{G}-\frac{\dot A}{A}\right)\right].
\label{acc}
\end{equation}
For the expansion scalar
\begin{eqnarray}
\Theta&=&V^\alpha_{;\alpha}\nonumber\\
&=&\frac{AB^2}{r^2A^2B^2+G^2}\,\left[r^2\left(2\frac{\dot B}{B}+\frac{\dot C}{C}\right)\right.\nonumber\\
&&+\left.\frac{G^2}{A^2B^2}\left(\frac{\dot B}{B}-\frac{\dot A}{A}+\frac{\dot G}{G}+\frac{\dot C}{C}\right)\right].
\label{theta}
\end{eqnarray}
Next, the shear tensor
\begin{equation}
\sigma_{\alpha \beta}= V_{(\alpha;\beta)}+a_{(\alpha}
V_{\beta)}-\frac{1}{3}\Theta h_{\alpha \beta}. \label{acc}
\end{equation}
where
\begin{equation}
h^\alpha_{\beta}=\delta ^\alpha_{\beta}+V^\alpha V_{\beta},
\label{vel5}
\end{equation}
 may be  defined through two scalar functions, as:

\begin{eqnarray}
\sigma _{\alpha \beta}=\frac{1}{3}(2\sigma _I+\sigma_{II}) (K_\alpha
K_\beta-\frac{1}{3}h_{\alpha \beta})\nonumber \\+\frac{1}{3}(2\sigma _{II}+\sigma_I) (L_\alpha
L_\beta-\frac{1}{3}h_{\alpha \beta}).\label{sigmaT}
\end{eqnarray}
The above scalars may be written in terms of the metric functions and their derivatives as (see \cite{1}):
\begin{eqnarray}
2\sigma _I+\sigma_{II}&=&\frac{3}{A}\left(\frac{\dot B}{B}-\frac{\dot C}{C}\right), \label{sigmasI}
\end{eqnarray}
\begin{eqnarray}
2\sigma _{II}+\sigma_I&=&\frac{3}{A^2B^2r^2+G^2}\,\left[AB^2r^2\left(\frac{\dot B}{B}-\frac{\dot C}{C}\right)\right.\nonumber\\
&
&\left.+\frac{G^2}{A}\left(-\frac{\dot A}{A}+\frac{\dot G}{G}-\frac{\dot C}{C}\right)\right] \label{sigmas},
\end{eqnarray}
 where the dot  and the prime denote derivatives with respect to $t$ and $r$ respectively. 
Once again, this specific choice of scalars, is justified by the very conspicuous way, in which they appear in the relevant equations (see the Appendix in \cite{1}).

Finally,  the vorticity  may be described, either by the vorticity vector  $\omega^\alpha$, or the vorticity tensor $\Omega
^{\beta\mu}$, defined as:
\begin{equation}
\omega_\alpha=\frac{1}{2}\,\eta_{\alpha\beta\mu\nu}\,V^{\beta;\mu}\,V^\nu=\frac{1}{2}\,\eta_{\alpha\beta\mu\nu}\,\Omega
^{\beta\mu}\,V^\nu,\label{vomega}
\end{equation}
where $\Omega_{\alpha\beta}=V_{[\alpha;\beta]}+a_{[\alpha}
V_{\beta]}$, and $\eta_{\alpha\beta\mu\nu}$ denote  the Levi-Civita tensor; we find a single component different from zero,  producing:

\begin{equation}
\Omega_{\alpha\beta}=\Omega (L_\alpha K_\beta -L_\beta
K_{\alpha}),\label{omegaT}
\end{equation}
and
\begin{equation}
\omega _\alpha =-\Omega S_\alpha.
\end{equation}
with the scalar function $\Omega$ given by
\begin{equation}
\Omega =\frac{G(\frac{G^\prime}{G}-\frac{2A^\prime}{A})}{2B\sqrt{A^2B^2r^2+G^2}}.
\label{no}
\end{equation}

Observe that from (\ref{no}) and regularity conditions at the centre, it follows that: $G=0\Leftrightarrow \Omega=0$.

Let us now introduce the electric ($E_{\alpha\beta}$) and magnetic ($H_{\alpha\beta}$) parts of the Weyl tensor ( $C_{\alpha \beta
\gamma\delta}$),  defined as usual by
\begin{eqnarray}
E_{\alpha \beta}&=&C_{\alpha\nu\beta\delta}V^\nu V^\delta,\nonumber\\
H_{\alpha\beta}&=&\frac{1}{2}\eta_{\alpha \nu \epsilon
\rho}C^{\quad \epsilon\rho}_{\beta \delta}V^\nu
V^\delta\,.\label{EH}
\end{eqnarray}

The electric part of the Weyl tensor has only three independent non-vanishing components, whereas only two components define the magnetic part. Thus  we may also write

\begin{widetext}
\begin{equation}
E_{\alpha\beta}=\frac{1}{3}(2\mathcal{E}_I+\mathcal{E}_{II}) (K_\alpha
K_\beta-\frac{1}{3}h_{\alpha \beta}) +\frac{1}{3}(2\mathcal{E}_{II}+\mathcal{E}_{I}) (L_\alpha
L_\beta-\frac{1}{3}h_{\alpha \beta})+\mathcal{E}_{KL} (K_\alpha
L_\beta+K_\beta L_\alpha), \label{E'}
\end{equation}
\end{widetext}
\noindent

and
\begin{equation}
H_{\alpha\beta}=H_1(S_\alpha K_\beta+S_\beta
K_\alpha)+H_2(S_\alpha L_\beta+S_\beta L_\alpha)\label{H'}.
\end{equation}

Also, for  the Riemann tensor we may define  three tensors $Y_{\alpha\beta}$, $X_{\alpha\beta}$ and
$Z_{\alpha\beta}$ as

\begin{equation}
Y_{\alpha \beta}=R_{\alpha \nu \beta \delta}V^\nu V^\delta,
\label{Y}
\end{equation}
\begin{equation}
X_{\alpha \beta}=\frac{1}{2}\eta_{\alpha\nu}^{\quad \epsilon
\rho}R^\star_{\epsilon \rho \beta \delta}V^\nu V^\delta,\label{X}
\end{equation}
and
\begin{equation}
Z_{\alpha\beta}=\frac{1}{2}\epsilon_{\alpha \epsilon \rho}R^{\quad
\epsilon\rho}_{ \delta \beta} V^\delta,\label{Z}
\end{equation}
 where $R^\star _{\alpha \beta \nu
\delta}=\frac{1}{2}\eta_{\epsilon\rho\nu\delta}R_{\alpha
\beta}^{\quad \epsilon \rho}$.

From the above tensor, we may define  the super--Poynting
vector  by
\begin{equation}
P_\alpha = \epsilon_{\alpha \beta \gamma}\left(Y^\gamma_\delta
Z^{\beta \delta} - X^\gamma_\delta Z^{\delta\beta}\right),
\label{SPdef}
\end{equation}
which, in our case  can be written as:
\begin{equation}
 P_\alpha=P_I K_\alpha+P_{II} L_\alpha,\label{SP}
\end{equation}

with
\begin{widetext}
\begin{eqnarray}
P_I =
\frac{2H_2}{3}(2{\cal E}_{II}+{\cal E}_I)+2H_1{\cal E}_{KL}+ \frac{32\pi^2 q_I}{3}\left[3(\mu+P)+\Pi_I\right] 
+32\pi^2 q_{II}\Pi_{KL} ,\nonumber
\\
P_{II}=-\frac{2H_1}{3}(2{\cal E}_{I}+{\cal E}_{II})-2H_2{\cal E}_{KL}
+ \frac{32\pi^2 q_{II}}{3}\left[3(\mu+P)+\Pi_{II}\right]+32\pi^2q_I\Pi_{KL} . \label{SPP}
\end{eqnarray}
\end{widetext}

In the theory of  the super--Poynting vector, a state of gravitational radiation is associated to a  non--vanishing component of the latter (see \cite{11p, 12p,  Basset, 14p}). This is in agreement with the established link between the super--Poynting vector and the news functions \cite{5p}, in the context of the Bondi--Sachs approach \cite{7, 8}. Furthermore, as it was shown in \cite{5p}, there is always a non-vanishing component of $P^\mu$, on the
plane orthogonal to a unit vector along which there is a non-vanishing component of vorticity (the $\theta$-$r$ plane).
Inversely, $P^\mu$ vanishes along the $\phi$-direction since there are no motions along this latter direction, because of the reflection symmetry. 

Therefore we can identify three different contributions in (\ref{SPP}). On the one hand we have contributions from the  heat transport process. These are in principle independent of the magnetic part of the Weyl tensor, which explains why they  remain in the spherically symmetric limit.  
  
On the other hand  we have contributions from the magnetic part of the Weyl tensor. It is reasonable to associate these with gravitational radiation. These are of two kinds. On the one hand contributions associated with the propagation of gravitational radiation within the fluid, and on the other, contributions of the flow of super--energy associated with the vorticity on the plane orthogonal to the direction of propagation of the radiation. Both contributions are intertwined, and it appears  to be impossible to disentangle them  through two independent scalars.

It is worth noticing   that the factors multiplying the $H$ terms in  (\ref{SPP}), are $\mathcal{E}_I,  \mathcal{E}_{II}, \mathcal{E}_{KL}$, implying that purely magnetic or purely electric sources, do not produce gravitational radiation. This is consistent with the result obtained in vacuum for the Bondi metric \cite{4p}, stating that purely electric Bondi metrics are static, whereas purely magnetic ones, are just Minkowski.

\section{The Kinematics}
 The following  discussion heavily relies on the kinematic quantities characterizing the motion of a medium presented in \cite{dem}, with slight changes in notation.

In Gaussian coordinates, the position of each particle may be given as
\begin{equation}
x^\alpha=x^\alpha(y^a,s),
\label{vel1}
\end{equation}
where $s$ is the proper time along the world line of the particle, and $y^a$ (with $a$ running from 1 to 3) is the position of the particle on a three-dimensional hypersurface (say $\Sigma$).
Then for the unit vector tangent to the world line (the four-velocity) we have
\begin{equation}
V^\alpha=\frac{\partial x^\alpha}{\partial s},
\label{vel2}
\end{equation}
and observe that
\begin{equation}
\frac{\partial }{\partial s}=D_T\equiv \frac{1}{A}\frac{\partial}{\partial t}.
\label{nueva}
\end{equation}
Next, for an infinitesimal variation of the world line we have
\begin{equation}
\delta x^\alpha=\frac{\partial x^\alpha}{\partial y^a} \delta y^a,
\label{vel3}
\end{equation}
from which it follows
\begin{equation}
D_T (\delta x^\alpha)=V^\alpha_{;\beta}\delta x^\beta.
\label{vel4b}
\end{equation}

We can define the position vector of the particle $y^a+\delta y^a$ relative to the particle $y^a$ on $\Sigma$, as
\begin{equation}
\delta_{\bot}x^\alpha=h^\alpha_{\beta} \delta x^\beta.
\label{vel6b}
\end{equation}
Then the relative velocity between these two particles, is
\begin{equation}
u^\alpha=h^\alpha_{\beta} D_T(\delta_{\bot} x^\beta),
\label{vel7b}
\end{equation}
and considering (\ref{vel4b}) and  (\ref{vel6b}) it follows that
\begin{equation}
u^\alpha=V^\alpha_{;\beta} \delta_{\bot} x^\beta.
\label{vel8}
\end{equation}

Now, the infinitesimal distance between two neighboring points on $\Sigma$ is
\begin{equation}
\delta l^2=g_{\alpha \beta} \delta_{\bot} x^\beta \delta_{\bot} x^\alpha,
\label{vel9}
\end{equation}
then
\begin{equation}
\delta l D_T(\delta l)=g_{\alpha \beta} \delta_{\bot} x^\beta D_T(\delta_{\bot} x^\alpha),
\label{vel10}
\end{equation}
or, by using (\ref{vel4b}) and (\ref{vel7b}),
\begin{equation}
\delta l D_T(\delta l)=V_{\alpha; \beta} \delta_{\bot} x^\beta \delta_{\bot} x^\alpha.
\label{vel11}
\end{equation}
Then, taking into consideration the expression for the irreducible components of a timelike vector
\begin{equation}
V_{\alpha
;\beta}=\sigma_{\alpha\beta}-a_{\alpha}V_{\beta}+\frac{1}{3}\Theta h_{\alpha\beta}+\Omega_{\alpha \beta},
\label{vel12}
\end{equation}
 
and, introducing the spacelike triad
\begin{equation}
e^\alpha_{(i)}\equiv (K^\alpha, L^\alpha, S^\alpha)=(\frac{\delta_{\bot} x^\alpha}{\delta l})_{(i)}, \label{unit}
\end{equation}
we obtain
\begin{equation}
 (\frac{D_T(\delta l)}{\delta l})_{(i,j)}= e^\alpha_{(i)} e^\beta_{(j )}(\sigma_{\alpha \beta}+\frac{\Theta}{3}h_{\alpha \beta}+\Omega_{\alpha \beta}),
\label{vel14}
\end{equation}
where triad indices $i,j$ run from $1$ to $3$.

From (\ref{vel14}) we can define the following  ``velocities'':
\begin{equation}
V_{(1)}=K^\alpha K^\beta(\sigma_{\alpha \beta}+\frac{1}{3}\Theta h_{\alpha \beta}+\Omega_{\alpha \beta}),
\label{vel1}
\end{equation}
\begin{equation}
V_{(2)}=L^\alpha L^\beta(\sigma_{\alpha \beta}+\frac{1}{3}\Theta h_{\alpha \beta}+\Omega_{\alpha \beta}),
\label{vel2}
\end{equation}
\begin{equation}
V_{(3)}=S^\alpha S^\beta(\sigma_{\alpha \beta}+\frac{1}{3}\Theta h_{\alpha \beta}+\Omega_{\alpha \beta}),
\label{vel3}
\end{equation}
\begin{equation}
V_{(1,2)}=K^\alpha L^\beta(\sigma_{\alpha \beta}+\frac{1}{3}\Theta h_{\alpha \beta}+\Omega_{\alpha \beta}),
\label{vel4}
\end{equation}
\begin{equation}
V_{(1,3)}=K^\alpha S^\beta(\sigma_{\alpha \beta}+\frac{1}{3}\Theta h_{\alpha \beta}+\Omega_{\alpha \beta}),
\label{vel5}
\end{equation}
which become, using (\ref{sigmaT}) and (\ref{omegaT})
\begin{equation}
V_{(1)}=\frac{1}{3}(\sigma_I+\Theta),\;
V_{(2)}=\frac{1}{3}(\sigma_{II}+\Theta),
\label{vel7}
\end{equation}
\begin{equation}
V_{(3)}=\frac{1}{3}(\Theta-\sigma_I-\sigma_{II}),\;
V_{(1,2)}=-\Omega,\;V_{(1,3)}=0,
\label{vel8}
\end{equation}
satisfying
\begin{equation}
V_{(1)}+V_{(2)}+V_{(3)}=\Theta.
\label{vel9}
\end{equation}

It is worth noticing  that the quantities defined above, describe variations of  $\delta l$, with respect to proper time,  (projected on different pairs of triad vectors), divided by $\delta l$. Accordingly, it would be perhaps more appropriate to call these quantities ``velocity contrast''  or ``specific velocities''.  However for simplicity we shall refer to them just as ``velocities''.

On the other hand, the geometrical and physical meaning of such quantities, becomes evident from (\ref{vel1})--(\ref{vel5}).

\section{The quasi--static regime}

Let us now translate the QSA defined in the first section, into conditions
to the different definitions of velocities given above,    and kinematical variables.

The fact that any characteristic time scale of the problem under consideration must be much smaller than the hydrostatic time,  implies that:
\begin{itemize}
\item The ``velocity'' functions $V_{(1),(2),(3)}$ and $V_{(1,2)}$ defined  in (\ref{vel1}-- \ref{vel4}) are small quantities (say of order $0(\epsilon)$, where $\epsilon<<1$).
\item We shall neglect all quantities of order $0(\epsilon^2)$ and higher.
\item From (\ref{vel7}, \ref{vel8}) it follows that $\sigma_{I, II}, \Theta, \Omega$, are of order $0(\epsilon)$.
\item From (\ref{theta}), (\ref{sigmasI}), (\ref{sigmas}), (\ref{no}) it follows that $\dot B$, $\dot C$  and $G$ are of order $0(\epsilon)$.
\item From (\ref{theta}), (\ref{sigmasI}), (\ref{sigmas}) it follows then, that up to the order $0(\epsilon)$ we have that $\sigma_I=\sigma_{II}\equiv \tilde \sigma$. And  up to the same order, 
\begin{equation}
\Theta+\tilde \sigma=\frac{3 \dot B}{AB}\quad \Theta-2\tilde \sigma=\frac{3\dot C}{AC}.
\label{fitem}
\end{equation}

\end{itemize}

Next, we have also to assume that the relaxation time in the transport equation (Eq.(57) in \cite{1}), must be neglected. 
Indeed, the relaxation time is 
the time taken by the system to return spontaneously to the steady state (whether of
thermodynamic equilibrium or not) after it has been suddenly removed from it. But as it follows from the very nature of the  QSA, all processes evolve on time scales which are much larger than the time scale on which transient phenomena take place, implying that we are assuming the heat flux vector to describe a steady heat flow.

Then, neglecting  the relaxation time in the transport equation, we obtain the following two equations (Eqs.(58,59) in \cite{1}, with $\tau=0$)

\begin{eqnarray}
q_{II}=-\frac{\kappa}{A}\left(\frac{G \dot T+A^2 T_{,\theta}}{ABr}+A T a_{II}\right) ,\label{qT1n}
\end{eqnarray}
and

\begin{eqnarray}
q_{I}=-\frac{\kappa}{B}(T^\prime+BTa_I)
. \label{qT2n}
\end{eqnarray}

Therefore, in the quasistatic regime, we obtain from the above equations, using  the fact that $\dot T$ is of order $0(\epsilon)$, and the conditions of thermal equilibrium,  \cite{Tolman}
\begin{equation}
(TA)^\prime=(TA)_{,\theta}=0,
\label{tol}
\end{equation}
that $\dot A$ is of order $0(\epsilon)$, whereas $\dot G$ is of order $0(\epsilon^2)$, which in turn implies  that up to the order $O(\epsilon)$:
\begin{equation}
a_I= \frac {A^\prime }{AB };\quad a_{II}=\frac {A_{,\theta}}{ABr}.
\label{accitem}
\end{equation}
From (\ref{no})) it follows at once that in our regime $\dot \Omega$ is of order $0(\epsilon^2)$ (a result that can also be obtained from B5 in \cite{1}).

From (B6, B7) in \cite{1}, in the QSA, we have respectively
\begin{equation}
\frac{2}{3B}\Theta^\prime-\frac{\Omega_{,\theta}}{Br}-\frac{\Omega}{Br}\left(\frac{2A_{,\theta}}{A}+\frac{C_{,\theta}}{C}\right)-\frac{\tilde\sigma^\prime}{3B}-\frac{\tilde \sigma C^\prime}{BC}=8\pi q_I,
\label{b6}
\end{equation}

\begin{equation}
\frac{2}{3Br}\Theta_{,\theta}+\frac{\Omega^\prime}{B}+\frac{\Omega}{B}\left(\frac{2A^\prime}{A}+\frac{C^\prime}{C}\right)-\frac{\tilde\sigma_{,\theta}}{3Br}-\frac{\tilde \sigma C_{,\theta}}{BCr}=8\pi q_{II}.
\label{b7}
\end{equation}
from which  it follows that dissipative fluxes are also of order $0(\epsilon)$.

Thus sumarizying all the consequences derived so far from the QSA we have:
\begin{itemize}
\item $\tilde\sigma, \Theta, \Omega$,  $\dot B$, $\dot C$, $\dot A$ , $q_I$, $q_{II}$, $\dot a_I$,  $\dot a_{II}$ and $G$ are of order $0(\epsilon)$.
\item $\dot \Omega$ is of order $0(\epsilon^2)$.
\end{itemize}

Next, from A6  in \cite{1}, in the QSA.
\begin{widetext}
\begin{eqnarray}
\frac{\dot \mu}{A} + (\mu+P)\Theta + \frac{\Pi_{I}}{9}\left(2\sigma_{I}+\sigma_{II}\right) + \frac{\Pi_{II}}{9}\left(2\sigma_{II}+\sigma_{I}\right)
+ \frac{q_{I}^\prime}{B} + \frac{1}{Br}\left(q_{II,\theta}+\frac{G}{A^2}\dot q_{II}\right)
 + 2q_{I}a_{I} 
+ 2q_{II}a_{II} \nonumber \\
+ \frac{q_{I}}{B}\left[\frac{C^\prime}{C}+\frac{(Br)^\prime}{Br}\right]
+ \frac{q_{II} }{Br} \left(\frac{B_{,\theta}}{B}+\frac{C_{,\theta}}{C}\right)=0\nonumber \\
\label{label}
\end{eqnarray}
\end{widetext}
it follows that $\dot \mu$ is of order $0(\epsilon)$.

From A7  in \cite{1}, we obtain the following two equations (in the QSA)
\begin{widetext}
\begin{eqnarray}
\frac{1}{B}\left(P+\frac{\Pi_{I}}{3}\right)^\prime+\frac{1}{Br}\left(\Pi_{KL,\theta}+\frac{G}{A^2}\dot \Pi_{KL}\right)
+\left(\mu+P+\frac{\Pi_{I}}{3}\right)a_{I}+\Pi_{KL}a_{II}\nonumber \\
+\frac{\Pi_{I}}{3B}\left[\frac{2C^\prime}{C}+\frac{(Br)^\prime}{Br}\right]%\nonumber \\
+\frac{\Pi_{II}}{3B}\left[\frac{C^\prime}{C}-\frac{(Br)^\prime}{Br}\right]
+ \frac{\Pi_{KL} }{Br} \left(\frac{2B_{,\theta}}{B}+\frac{C_{,\theta}}{C}\right)
+\frac{\dot q_{I}}{A}=0
\label{label1}
\end{eqnarray}
\end{widetext}
and
\begin{widetext}
\begin{eqnarray}
\frac{1}{Br}\left[\left(P+\frac{\Pi_{I}}{3}\right)_{,\theta}+\frac{G}{A^2}\left(\dot P +\frac{\dot \Pi_{II}}{3}\right)\right]+\frac{\Pi_{KL}^\prime}{B}
+\left(\mu+P+\frac{\Pi_{II}}{3}\right)a_{II}+\Pi_{KL}a_{I}\nonumber \\
+ \frac{\Pi_{I}}{3Br} \left(-\frac{B_{,\theta}}{B}+\frac{C_{,\theta}}{C}\right)
+ \frac{\Pi_{II} }{3Br} \left(\frac{B_{,\theta}}{B}+\frac{2C_{,\theta}}{C}\right)
+\frac{\Pi_{KL}}{B}\left[\frac{C^\prime}{C}+2\frac{(Br)^\prime}{Br}\right]
+\frac{\dot q_{II}}{A}=0
\label{label2}
\end{eqnarray}
\end{widetext}
Since the hydrostatic equilibrium condition holds  at any time, the corresponding hydrostatic equilibrium equations (Eqs. (21,22) in \cite{static}) must be satisfied. Then, from the two equations above, we obtain respectively
\begin{equation}
\dot \Pi_{KL}\approx 0(\epsilon);\quad \dot q_I \approx 0(\epsilon^2);\quad \ddot B\approx 0(\epsilon^2)\quad \ddot C\approx 0(\epsilon^2).
\label{hydro1}
\end{equation}
and
\begin{equation}
\dot \Pi_{II}\approx 0(\epsilon);\quad \dot q_{II} \approx 0(\epsilon^2)\quad \dot P\approx 0(\epsilon),
\label{hydro11}
\end{equation}
where the fact has been used that in the QSA, $P, \Pi_I, \Pi_{II}$ contain (besides terms including spatial derivatives of the metric tensor), terms with $\ddot B$ and $\ \ddot C$. 

From (\ref{hydro1}) and (\ref{fitem}) it follows at once that
\begin{equation}
\dot{\tilde\sigma}\approx 0(\epsilon^2);\quad \dot \Theta\approx 0(\epsilon^2).
\label{hydro17}
\end{equation}

Let us now turn to  (\ref{b6}), which, using (\ref{vel7}) and (\ref{vel8}), may be written as:
\begin{eqnarray}
2V^\prime=\tilde \sigma\left[\ln{(\tilde \sigma C)}\right]^\prime+ \frac{\Omega}{r}\left[\ln{(\Omega A^2 C)}\right]_{,\theta} +8\pi q_I B,
\label{b8}
\end{eqnarray}
with $V\equiv V_{(1)}\equiv V_{(2)}$.

After integration we obtain
\begin{equation}
V=V_\Sigma-\frac{1}{2}\int^{r_\Sigma}_r\{\tilde \sigma\left[\ln{(\tilde \sigma C)}\right]^\prime+ \frac{\Omega}{r}\left[\ln{(\Omega A^2 C)}\right]_{,\theta} +8\pi q_I B\}dr,
\label{b9}
\end{equation}
or
\begin{widetext}
\begin{equation}
V_{(3)}=V_{(3)\Sigma}-\frac{1}{2}\int^{r_\Sigma}_r\left\{\tilde \sigma\left[\ln({\frac{C}{\tilde \sigma}})\right]^\prime+ \frac{\Omega}{r}\left[\ln{(\Omega A^2 C)}\right]_{,\theta} +8\pi q_I B\right\}dr,
\label{b10}
\end{equation}
\end{widetext}
where the boundary surface of the source is defined by the equation $r=r_{\Sigma}$ and the fact that $V_{(3)}=V-\tilde \sigma$, has been used.

In a similar way  we may writte down (\ref{b7}) as
\begin{equation}
2V_{,\theta}=\tilde \sigma\left[\ln{(\tilde \sigma C)}\right]_{,\theta}-\Omega r\left[\ln{\left(\frac{ A^2 C}{\Omega}\right)}\right]^\prime +8\pi q_{II} Br,
\label{b11}
\end{equation}
producing
\begin{equation}
V=V_\Sigma-\frac{1}{2}\int^{\theta_\Sigma}_\theta\left\{\tilde \sigma\left[\ln{(\tilde \sigma C)}\right]_{,\theta}-r\Omega\left[\ln{\left(A^2 C \Omega\right)}\right]^\prime +8\pi q_{II} Br\right\}d\theta,
\label{b12}
\end{equation}
or
\begin{widetext}
\begin{equation}
V_{(3)}=V_{(3)\Sigma}-\frac{1}{2}\int^{\theta_\Sigma}_\theta\left\{\tilde \sigma\left[\ln{\frac{C}{\tilde \sigma }}\right]_{,\theta}-r\Omega\left[\ln{\left(A^2 C \Omega\right)}\right]^\prime +8\pi q_{II} Br\right\}d\theta,
\label{b13}
\end{equation}
\end{widetext}
where now the boundary surface equation is given by $\theta=\theta_\Sigma$.

Let us focus on the expressions above (\ref{b9}), (\ref{b10}), (\ref{b12}), (\ref{b13}). If we assume the fluid to be irrotational, shear--free and dissipationless, then the sign of $V$ and  $V_{(3)}$ is the same as the sign of $V_{\Sigma}$   and $V_{(3)\Sigma}$, for any fluid element within the distribution. However, the presence of any of the factors  above (vorticity, shear, heat flux), may lead to  a situation, where either velocity changes of sign within the fluid distribution, with respect to its sign on the boundary surface. In other words, it may happen that some inner regions move in one direction whereas the outer ones move in the opposite direction. Such  ``splittings'' of the configuration, have already been reported for the spherically symmetic case (see \cite{two}, \cite{split1}, \cite{split}, \cite{p1}, \cite{p2} and references therein). Here the picture is more involved than in the spherically symmetric case due to the possibility of the splitting to occur, along  two orthogonal directions.

Next, Eqs. (B8, B9) in \cite{1},  read in the QSA as
\begin{equation}
H_1=-\Omega a_I-\frac{1}{2B}\left(\Omega^\prime-\frac{\Omega C^\prime}{C}\right)-\frac{1}{2Br}\left(\tilde \sigma_{,\theta}+\frac{\tilde \sigma C_{,\theta}}{C}\right),
\label{H1}
\end{equation}
\begin{equation}
H_2=-\Omega a_{II}-\frac{1}{2Br}\left(\Omega_{,\theta}-\frac{\Omega C_{,\theta}}{C}\right)+\frac{1}{2B}\left(\tilde \sigma^\prime+\frac{\tilde \sigma C^\prime}{C}\right),
\label{H2}
\end{equation}
implying that the magnetic part of the Weyl tensor is of order $0(\epsilon)$.

It is worth noticing that in the vorticity--free case( $G=\Omega=0$), it follows from (\ref{H1}) and (\ref{H2})  that

\begin{equation}
H_1=-\frac{(\tilde \sigma C)_{,\theta}}{2BrC},\qquad  H_2=\frac{(\tilde \sigma C)^\prime}{2BC}.
\label{vf1}
\end{equation}

Then, from (\ref{b6}) and (\ref{b7}), using (\ref{fitem}), we find (always assuming $\Omega=0$)
\begin{equation}
2\left(\frac{\dot B}{AB}\right)^\prime=\frac{(\tilde \sigma C)^\prime}{C}+8\pi Bq_I,
\label{vf2}
\end{equation}

and 
\begin{equation}
2\left(\frac{\dot B}{AB}\right)_{,\theta}=\frac{(\tilde \sigma C)_{,\theta}}{C}+8\pi Brq_{II}.
\label{vf3}
\end{equation}

The combination of the two equations above with (\ref{vf1}) produces
\begin{equation}
H_1=-\frac{1}{Br}\left(\frac{\dot B}{AB}\right)_{,\theta}+4\pi q_{II},
\label{vf4}
\end{equation}
and
\begin{equation}
H_2=\frac{1}{B}\left(\frac{\dot B}{AB}\right)^\prime-4\pi q_{I}.
\label{vf5}
\end{equation}

Thus from (\ref{vf1}) it follows at once that, in the QSA, for the vorticity--free case, the vanishing of the shear is a necessary and sufficient condition for the fluid to be purely electric.  This result is somehow complementary to the one obtained in the general dynamic case \cite{HDO}, which states   that, for a dissipative and anisotropic, shear--free  fluid, the  vanishing vorticity, is a  necessary and sufficient condition for the magnetic part of the Weyl tensor  to vanish, this last result in turn provides a generalization of  a similar result for perfect fluids, obtained in  \cite{b1, b2, glass}.
\section{Conclusions}
Wew have provided a general framework for describing the evolution of axially symmetric dissipative fluids in the QSA. 

The role played by  the vorticity, the shear  and the dissipative flux is clearly brought out, through the expressions (\ref{b9}), (\ref{b10}) and (\ref{b12}), (\ref{b13}). Such expressions show how the fluid distribution may split, under the effects of the factors mentioned above, leading  to a variety of very different structures.

Finally it is worth mentioning that in the QSA the magnetic part of the Weyl tensor does not necessarily vanish (though it is of order $O(\epsilon)$), thereby implying that the ``gravitational'' part of the super-Poynting vector does not vanish either, meaning that gravitational radiation is not incompatible with the QSA. 

 However, from (\ref{H1}) and (\ref{H2}) (alternatively see (\ref{B17}) and (\ref{B18})), it follows at once that $\dot H_1\approx \dot H_2 \approx O(\epsilon^2)$ (at least), and therefore are neglectable in the QSA. This in turn implies that if the magnetic part of the Weyl tensor vanishes at any given time, it will do so for any time afterwards. In other words, no state of radiation for a finite period of time is expected in the QSA. This result is in agreement with the one obtained by Bondi \cite{Bn}, for a more restricted case. However besides the ``inductive'' transfer of energy, mentioned by Bondi, we also have here, the transfer carried on by the dissipative flux.
\\

\begin{acknowledgments}
L.H. thanks  Departament de F\'isica at the  Universitat de les  Illes Balears, for financial support and hospitality. ADP  acknowledges hospitality of the
 Departament de F\'isica at the  Universitat de les  Illes Balears. J.O. acknowledges financial support from the Spanish
Ministry of Science and Innovation (grant FIS2009-07238).
\end{acknowledgments}

\appendix* 
\section{The remaining equations in the QSA}
Below, for the benefit of the reader, we shall write  the equations not used explicitly in the text,  specialized  for the QSA, from  the framework developped in \cite{1}.  Thus, Eqs. (B10--B18) in \cite{1}, read respectively

\begin{widetext}

\begin{eqnarray}
\frac{1}{3A} \left({\cal E}_{I} +4\pi \Pi_{I} +4\pi \mu\right)^{.} +\frac{1}{3}\left({\cal E}_{I} \Theta + {\cal E}_{II} \tilde\sigma\right) - \Omega \left({\cal E}_{KL} + 4\pi \Pi_{KL}\right)
- \frac{1}{Br}\left(H_{1,\theta}+H_1 \frac{ C_{,\theta}}{C} \right) +\frac{H_2}{B}\left(\frac{(Br)^\prime}{Br}- \frac{ C^\prime}{C}  \right)\nonumber \\= -\frac{4\pi}{3}\left(\mu + P + \frac{\Pi_{I}}{3}\right)\left(\tilde \sigma + \Theta\right)+2a_{II}H_1-8\pi a_{I}q_{I}-\frac{4 \pi }{B}\left(q^\prime_{I}+ q_{II} \frac{B_{,\theta}}{Br}\right)\nonumber\\
\label{B10}
\end{eqnarray}

\begin{eqnarray}
\frac{1}{A} \left({\cal E}_{KL} +4\pi \Pi_{KL}\right)^{.} +\frac{1}{6}\Omega \left[{\cal E}_{I} - {\cal E}_{II} +4\pi \left(\Pi_{I}-\Pi_{II}\right) \right]  + \left({\cal E}_{KL} + 4\pi \Pi_{KL}\right)\left(\Theta - \tilde\sigma\right) +a_{I} H_{1} - a_{II }H_{2}\nonumber\\
+\frac{1}{2B}\left[H_1^\prime - H_1 \left(\frac{(Br)^\prime}{Br}-\frac{2C^\prime}{C}\right)\right]+\frac{1}{2Br}\left[-H_{2,\theta} + H_2 \left(\frac{B_{,\theta}}{B}-\frac{2C_{,\theta}}{C}\right)\right]=\frac{8\pi \Pi_{KL}}{3}\left(\Theta-2\tilde \sigma\right)\nonumber \\
-4\pi\left(a_{II}q_{I}+a_{I}q_{II}\right)-\frac{2\pi}{B}\left(q_{II}^\prime-q_{II}\frac{(Br)^\prime}{Br}\right) -\frac{2\pi}{Br}\left(q_{I,\theta}-q_{I}\frac{B_{,\theta}}{B}\right)
\label{B11}
\end{eqnarray}

\begin{eqnarray}
\frac{1}{3A} \left[4\pi\left(\mu+ \Pi_{II}\right)+ {\cal E}_{II}\right]^{.} +\frac{1}{3}\left({\cal E}_{II} \Theta + {\cal E}_{I} \tilde\sigma\right) + \Omega \left({\cal E}_{KL} + 4\pi \Pi_{KL}\right)+2a_{I} H_{2}  +\frac{1}{B}\left(H_2^\prime +H_2 \frac{C^\prime}{C}\right)\nonumber \\
-\frac{H_1}{Br}\left(\frac{B_{,\theta}}{B}-\frac{C_{,\theta}}{C}\right)
= -\frac{4\pi}{3}\left(\mu + P + \frac{\Pi_{II}}{3}\right)\left(\tilde \sigma + \Theta\right)-8\pi a_{II}q_{II}-\frac{4 \pi q_{II,\theta}}{Br}-\frac{4\pi q_{I}}{B}\frac{(Br)^\prime}{Br}
\label{B12}
\end{eqnarray}

\begin{eqnarray}
\frac{1}{3A}\left[4\pi\left(\mu-\Pi_{I}-\Pi_{II}\right)-\left({\cal E}_{I}+{\cal E}_{II}\right)\right]^{.}-\frac{4\pi}{9}\left(\Pi_{I}+\Pi_{II}\right)\left(\Theta-2\tilde \sigma\right)
-\frac{1}{3}\left({\cal E}_{I}+{\cal E}_{II}\right)\left(\Theta+\tilde\sigma\right)\nonumber \\
+2\left(H_1 a_{II}-H_2 a_{I}\right)-\frac{1}{B}\left(H_2^\prime + H_2 \frac{(Br)^\prime}{Br}\right)+\frac{1}{Br}\left(H_{1,\theta}+H_1 \frac{B_{,\theta}}{B}\right)\nonumber\\
=\frac{4\pi}{3}(\mu+P)\left(2\tilde \sigma -\Theta\right)
-\frac{4 \pi q_{I}}{B}\frac{C^\prime}{C}-\frac{4 \pi q_{II}}{Br}\frac{C_{,\theta}}{C}
\label{B13}
\end{eqnarray}

\begin{eqnarray}
\frac{1}{3B}\left({\cal E}_{I}+4\pi \Pi_{I}\right)^\prime+\frac{1}{Br}\left({\cal E}_{KL}+4\pi \Pi_{KL}\right)_{,\theta}+\frac{1}{3B}\left({\cal E}_{I}+4\pi \Pi_{I}\right)\left(\frac{(Br)^\prime}{Br}+\frac{2C^\prime}{C}\right)
\nonumber\\ 
-\frac{1}{3B}\left({\cal E}_{II}+4\pi \Pi_{II}\right)\left(\frac{(Br)^\prime}{Br}-\frac{C^\prime}{C}\right)+\frac{1}{Br}\left({\cal E}_{KL}+4\pi \Pi_{KL}\right)\left(\frac{2B_{,\theta}}{B}+\frac{C_{,\theta}}{C}\right)=\frac{8\pi}{3B}\mu^\prime
\label{B14}
\end{eqnarray}

\begin{eqnarray}
\frac{1}{3Br}\left({\cal E}_{II}+4\pi \Pi_{II}\right)_{,\theta}+\frac{1}{B}\left({\cal E}_{KL}+4\pi \Pi_{KL}\right)^\prime-\frac{1}{3Br}\left({\cal E}_{I}+4\pi \Pi_{I}\right)\left(\frac{B_{,\theta}}{B}-\frac{C_{,\theta}}{C}\right)
\nonumber\\ 
+\frac{1}{3Br}\left({\cal E}_{II}+4\pi \Pi_{II}\right)\left(\frac{B_{,\theta}}{B}+\frac{2C_{,\theta}}{C}\right)+\frac{1}{B}\left({\cal E}_{KL}+4\pi \Pi_{KL}\right)\left(\frac{2(Br)^\prime}{Br}+\frac{C^\prime}{C}\right)=\frac{8\pi}{3Br}\mu_{,\theta}
\label{B15}
\end{eqnarray}

\begin{eqnarray}
-\frac{1}{B}\left[H_1^\prime +H_1\left(\frac{(Br)^\prime}{Br}+\frac{2C^\prime}{C}\right)\right]-\frac{1}{Br}\left[H_{2,\theta} +H_2\left(\frac{B_{,\theta}}{B}+\frac{2C_{,\theta}}{C}\right)\right]=\nonumber \\
\left[8\pi\left(\mu+P\right)+\frac{4\pi}{3}\left(\Pi_{I}+\Pi_{II}\right)-\left({\cal E}_{I}+{\cal E}_{II}\right)\right] \Omega-\frac{4\pi}{Br}\left(q_{I,\theta}+q_{I}\frac{B_{,\theta}}{B}\right)+\frac{4\pi}{B}\left(q_{II}^\prime+q_{II}\frac{(Br)^\prime}{Br}\right)
\label{B16}
\end{eqnarray}

\begin{eqnarray}
-\frac{4\pi}{B}\Pi_{KL}^\prime-\frac{1}{3Br}\left({\cal E}_{I}+{\cal E}_{II}-4\pi\Pi_{I}\right)_{,\theta}+\frac{{\cal E}_{KL}}{B}\left(\frac{A^\prime}{A}-\frac{C^\prime}{C}\right)-\frac{8\pi\Pi_{KL}}{B}\frac{(Br)^\prime}{Br}\nonumber\\
 -\frac{{\cal E}_{I}}{3Br}\left(\frac{2A_{,\theta}}{A}+\frac{C_{,\theta}}{C}\right)
-\frac{{\cal E}_{II}}{3Br}\left(\frac{A_{,\theta}}{A}+\frac{2C_{,\theta}}{C}\right)+\frac{4\pi}{3Br}\left(\Pi_{I}-\Pi_{II}\right)\frac{B_{,\theta}}{B}+\frac{\dot H_1}{A}=-\frac{4\pi}{3Br}\mu_{,\theta}
\label{B17}
\end{eqnarray}

\begin{eqnarray}
\frac{1}{3B}\left({\cal E}_{I}+{\cal E}_{II}-4\pi\Pi_{II}\right)^\prime+\frac{4\pi}{Br}\left(\Pi_{KL,\theta}+\Pi_{KL}\frac{2B_{,\theta}}{B}\right)+\frac{{\cal E}_{I}}{3B}\left(\frac{A^\prime}{A}+\frac{2C^\prime}{C}\right)+\frac{{\cal E}_{II}}{3B}\left(\frac{2A^\prime}{A}+\frac{C^\prime}{C}\right)\nonumber \\
-\frac{{\cal E}_{KL}}{Br}\left(\frac{A_{,\theta}}{A}-\frac{C_{,\theta}}{C}\right)
+\frac{4\pi}{3B}\left(\Pi_{I}-\Pi_{II}\right)\frac{(Br)^\prime}{Br}+\frac{\dot H_2}{A}=\frac{4\pi}{3B}\mu^\prime
\label{B18}
\end{eqnarray}

\end{widetext}

\end{document}